\documentstyle[psfig]{lamuphys}
\makeatletter
\let\chapter\hid@chapter
\makeatother
\begin{document}

\pagenumbering{arabic}

\title{Interpolative method for transport properties of quantum dots in
the Kondo regime} 

\author{A. Levy Yeyati, A. Mart\'{\i}n-Rodero and F. Flores}

\institute{Departamento de F\'{\i}sica Te\'orica de la Materia
Condensada CV, Universidad Autonoma de Madrid, 28049 Madrid, Spain}

\titlerunning{Tranport properties of quantum dots in the Kondo regime}

\maketitle

\begin{abstract} 
We present an interpolative method for describing coherent transport
through an interacting quantum dot. The idea of the method is to
construct an approximate electron self-energy which becomes exact both
in the limits of weak and strong coupling to the leads. The validity of
the approximation is first checked for the case of a single 
(spin-degenerate) dot level. A generalization to the multilevel case is
then discussed. We present results both for the density of states and the
temperature dependent linear conductance showing the transition from the
Kondo to the Coulomb blockade regime.    
\end{abstract}
\section{Introduction}
The Kondo effect constitutes a prototypical correlation effect in
condensed matter physics. Although originally studied in connection to
magnetic impurities in metals, there is now a renewed interest in this
many-body problem fostered by the recent observation of Kondo effect in
semiconducting quantum dots \cite{Goldhaber}. Quantum dots
provide an almost ideal laboratory where the relevant parameters can be
controlled, which allow to test the predictions of theoretical
models.

From the theoretical side, Kondo physics in quantum dots has been mainly
analyzed in the light of the so called single level Anderson model.
There were predictions for the Kondo effect in quantum dots based on
this model since the early 90's \cite{Ng}. 
The theory predicts an enhancement of the linear conductance due to
Kondo effect at very low temperatures, which is in qualitative agreement
with recent experiments.

However, in most realistic situations, the single level Anderson model
constitutes a crude approximation for a quantum dot. Actual semiconducting
quantum dots contain a large number ($\sim 100$) of electrons and the
single-particle level separation between dot levels may be not so large 
compared to the level broadening, which restricts the validity of the
single-level approximation. The actual situation would be more
appropriately described by a multilevel model, including several instead
of a single dot level. Unfortunately, there are no simple theoretical
approaches to extract the electronic and transport properties from such
a microscopic model. 

In this paper we present results on the Kondo effect in quantum dots
based on the interpolative method.  
The basic idea of this method is to construct an interpolative electron
self-energy which becomes exact both in the limits of weak and strong
coupling to the leads. These ideas were first introduced in Ref.
\cite{Alvaro} in connection to the single-level Anderson model and have, 
since then,
been adapted by several authors to different problems involving strongly
correlated electrons. In this way, the method has been used to study the
Hubbard model \cite{AlvaroII}, the non-equilibrium Anderson model
\cite{us}, the metal-insulator transition in infinite dimensions
\cite{Kajueter}, to incorporate correlation effects into band-structure
calculations \cite{LK}, the ac-Kondo effect in quantum dots \cite{CT}
and finally extended by the present authors to analyze 
the multilevel Anderson model \cite{usII}. 

The paper will be organized as follows: In section 2 we present the
interpolative method. We first discuss the single level case, showing
the accuracy of the method with the help of a simple exactly solvable
model. We then consider the multilevel situation. In section 3 we
present results which illustrate the behavior of the conductance with
temperature in a multilevel situation. 

\section{The interpolative method}

For describing a multilevel quantum dot (QD) we consider a model Hamiltonian
$H= H_{\mbox{dot}} + H_{\mbox{leads}} + H_T$ 
where $H_{\mbox{dot}} = \sum_{m} \epsilon_m
\hat{d}^{\dagger}_m \hat{d}_m + U \sum_{l > m} \hat{n}_m \hat{n}_l$ 
corresponds to the uncoupled QD ($\hat{n}_m = \hat{d}^{\dagger}_m \hat{d}_m$); 
$H_{\mbox{leads}} = \sum_{k \in L,R} \epsilon_k \hat{c}^{\dagger}_k \hat{c}_k$ 
to the uncoupled leads, and $H_T = \sum_{m, k \in L,R}
t_{m,k} \hat{d}^{\dagger}_m \hat{c}_{k} + h.c.$ describes the coupling
between the dot and the leads. The labels $m$ and $l$ in $H$ denote the 
different dot levels including spin quantum numbers. The number of dot
levels will be denoted by $M$ (i.e. $1\le m,l \le M$). We adopt the usual
simplifying assumption of having the same electron-electron interaction
$U$ between any pair of dot states.

The main objective of our method is to determine the dot retarded Green
functions $G_m(\tau) = -i \theta(\tau) 
< [\hat{d}_m(\tau),\hat{d}^{\dagger}_m(0)]_+> $ from which the different
level charges and the dot linear conductance can be obtained. In the
frequency representation we can write $G_m$ as:
\begin{equation} \label{retarded}
G_m(\omega) =  \frac{1}{\omega - \epsilon^{HF}_m  - \Sigma_m(\omega) 
- \Gamma_{m,L}(\omega) - \Gamma_{m,R}(\omega)}, 
\end{equation} 
where $\epsilon^{HF}_m = \epsilon_m + U \sum_{l \ne m} n_l$ is the
Hartree-Fock level (we adopt the notation $n_l$ for the mean charge on
level $l$) and $\Gamma_{m,L},\Gamma_{m,R}$ are tunneling rates coupling
the dot to the leads, given by $\Gamma_{m,L(R)}(\omega) = \sum_{k\in L(R)} 
t_{m,k}^2/(\omega - \epsilon_k + i0^+)$.
We shall neglect indirect coupling between dot levels through the leads
(non-diagonal elements $\Gamma_{m,m^{\prime},L(R)}$) and adopt the usual
approximation of considering $\Gamma_{m,L(R)}$ as a pure imaginary
constant independent of the energy.

The self-energy $\Sigma_m(\omega)$ takes into account electron
correlation effects beyond the Hartree approximation. The idea of the
present approximation is to determine an interpolative self-energy which
yields the correct exact results both in the $\Gamma/U \rightarrow 0$
limit (atomic limit) and in the opposite $U/\Gamma \rightarrow 0$ limit.

\subsection{The single-level case}

Let us first discuss how to proceed for the simple single-level case.
In this case $m=1,2$, the two indexes corresponding to up and down spin
orientations. These will be denoted by $\sigma$ and $\bar{\sigma}$. 
In the atomic limit $G_{\sigma}$ can be obtained using the equation of
motion technique \cite{Bell} as
\begin{equation} \label{atomicsl}
G^{(at)}_{\sigma}(\omega) =  
\frac{1 - n_{\bar{\sigma}}}{\omega - \epsilon + i0^+} + 
\frac{n_{\bar{\sigma}}}{\omega - \epsilon - U + i0^+}  
\end{equation} 

This expression can be formally written in the usual Fermi liquid form,
i.e. $G^{(at)}_{\sigma}(\omega) = [\omega - \epsilon - U n_{\bar{\sigma}}
- \Sigma^{(at)}_{\sigma}(\omega)]^{-1}$ by introducing the ``atomic''
self-energy 

\[\Sigma^{(at)}_{\sigma}(\omega) = \frac{U^2 n_{\bar{\sigma}} (1 -
n_{\bar{\sigma}})}{\omega - \epsilon - U(1 - n_{\bar{\sigma}}) + i0^+} \]

In the opposite limit, $U/\Gamma \rightarrow 0$, the electron self-energy
can be calculated by second-order perturbation theory in $U$, which
yields
\begin{eqnarray}
\Sigma^{(2)}_{\sigma}(\omega) & = & U^2 
\int_{-\infty}^{\infty} d\epsilon_1 \int_{-\infty}^{\infty} d\epsilon_2
\int_{-\infty}^{\infty} d\epsilon_3 \; 
\frac{\tilde{\rho}_{\sigma}(\epsilon_1) 
\tilde{\rho}_{\bar{\sigma}}(\epsilon_2) 
\tilde{\rho}_{\bar{\sigma}}(\epsilon_3)}{\omega + \epsilon_2 - \epsilon_1 - 
\epsilon_3 + i0^+} \times \nonumber \\
& & \left[ f(\epsilon_1) f(\epsilon_3) 
\left(1 - f(\epsilon_2) \right) + \left(1 - f(\epsilon_1) \right) 
\left(1 - f(\epsilon_3)\right) f(\epsilon_2) \right],
\end{eqnarray}
where $f(\omega)$ is the Fermi distribution function and
$\tilde{\rho}_{\sigma}(\omega) = 
\Gamma/\pi((\omega-\tilde{\epsilon}_{\sigma})^2 + \Gamma^2)$ 
is the local density of states for an effective level
$\tilde{\epsilon}_{\bar{\sigma}}$, which will be determined in order to 
fulfill exact Fermi liquid properties at zero temperature. 

It is important to stress the following simple property of
$\Sigma^{(2)}$: 

\[ \lim_{\Gamma \rightarrow 0} \Sigma^{(2)}_{\sigma}(\omega) \;\; = U^2
\frac{\tilde{n}_{\bar{\sigma}}(1 - \tilde{n}_{\bar{\sigma}})}{\omega -
\tilde{\epsilon}_{\sigma} + i0^+} \]  

Thus, when extrapolated to the atomic limit $\Sigma^{(2)}$ has the same
functional form as $\Sigma^{(at)}$. This property suggests that one can
smoothly interpolate between the two limits. The ansazt proposed in Ref.
\cite{Alvaro} for the interpolative self-energy is:

\begin{equation}
\Sigma_{\sigma}(\omega) = 
\frac{\Sigma^{(2)}_{\sigma}(\omega)}{1 - \alpha \; 
\Sigma^{(2)}_{\sigma}(\omega)}
\end{equation} 
where $\alpha = (\epsilon - \tilde{\epsilon}_{\sigma} - U 
(1 - n_{\bar{\sigma}}))/(U^2 n_{\bar{\sigma}} (1 - n_{\bar{\sigma}}))$.
This ansazt has the desired property $\Sigma \rightarrow \Sigma^{(2)}$ when $U
\rightarrow 0$ and $\Sigma \rightarrow \Sigma^{(at)}$ when $\Gamma
\rightarrow 0$. 

The final step is to impose the proper self-consistent condition for
determining the effective level $\tilde{\epsilon}$. At zero temperature,
from the Luttinger-Ward relations \cite{LW} one can derive the Friedel
sum rule for the Anderson model \cite{Langreth} 
\[ n_{\sigma} = -\frac{1}{\pi} \mbox{Im} \ln G^r_{\sigma}(E_F) \]
which imposes an exact relation between the dot-level charge and the phase
shift at the Fermi energy. The effective level can thus be determined 
in order to fulfill the Friedel sum rule. 
This condition is, however, not valid at finite temperature. In Ref.
\cite{us} we show that the condition $n_{\sigma} = \tilde{n}_{\sigma}$,
i.e. imposing the same charge in the effective system as in the
interacting system, is approximately equivalent to the Friedel sum rule 
at zero temperature but can be also used at finite temperature.

In order to check the accuracy of the interpolative method we have
considered a simple two-sites problem that can be diagonalized exactly.
One of the sites would describe the metallic leads and the other site
corresponds to the dot. In order to analyze the more general situation
we impose a finite splitting $\Delta = \epsilon_{\sigma} -
\epsilon_{\bar{\sigma}}$ between the two spin orientations on the dot. 
Within this toy model the second order self-energy can be evaluated
analytically.

In figure 1 we show the charge on the two dot levels as a function of
gate voltage (the gate voltage is the distance between the lower dot
level and the leads level). As can be
observed, in the exact solution there is a blocking of the upper level
charge until the gate voltage becomes larger than $\Delta + U$. The
exact behavior is accurately reproduced by the interpolative method. 
It is instructive to consider another simple approximation 
widely used in the literature, which consist in just broadening the
poles in the atomic Green function (2) by the non-interacting tunneling
rates. This approximation corresponds to the so-called Hubbard I approximation
\cite{Hubbard}. As can be observed in the lower panel of Fig. 1, this
approximation fails to give the blocking of the upper level found in the
exact solution.

\begin{figure}[thbp]
\psfig{file=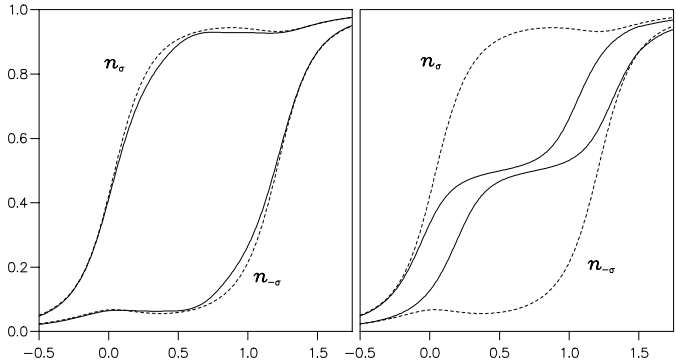,height=4cm}
\caption[]{Level charges as a function of gate voltage for the two sites
model with $\Delta = 0.25$ and $t=0.1$ (in units of the charging energy
$U$). Left pannel corresponds to the interpolative approach and
the right pannel to the Hubbard I approach. The exact solution is shown
as a dashed line.}
\end{figure}

\subsection{Multilevel case}

The multilevel version of the interpolative method is somewhat more
complex \cite{usII}. In the first place, the atomic limit Green functions 
do not contain just two poles but several poles corresponding to the various 
different charge states of the dot. The corresponding expression can be
obtained using the equation of motion technique and is given by

\begin{eqnarray}
G^{(at)}_m(\omega) & = & \frac{<\prod_{l \ne m} (1 - \hat{n}_l)>}{\omega -
\epsilon_m + i0^+} + \sum_{l \ne m} \frac{<\hat{n}_l 
\prod_{(s \ne l) \ne m} (1 -
\hat{n}_s)>}{\omega - \epsilon_m - U + i0^+} + ...  \nonumber \\
& & + \frac{<\prod_{l \ne m} \hat{n}_l>}{\omega - \epsilon_m - (M-1)U + i0^+} ,
\end{eqnarray}

The evaluation of this expression requires the knowledge of up to
$M-1$-body correlations functions $<\hat{n}_1 \hat{n}_2>$, 
$<\hat{n}_1 \hat{n}_2 \hat{n}_3>$, ..., etc. However, for sufficiently
large $U$ fluctuactions in the dot charge by more of one electron with
respect to the mean charge $\cal{N}$ are strongly inhibited. One can
thus approximate Eq. (5) as follows

\begin{eqnarray}
G^{(at)}_m(\omega) & \simeq &
\frac{A^m_{N-1}}{\omega - \epsilon_m - U(N-1) + i0^+}
+ \frac{A^m_N}{\omega - \epsilon_m  - U N + i0^+} \nonumber \\
& & + \frac{A^m_{N+1}}{\omega - \epsilon_m  - U(N+1) + i0^+}, 
\end{eqnarray} 
where $N = Int[\cal{N}]$. In order to yield the exact first three
momenta of the exact spectral density the weight factors $A^m_N$ should satisfy 
the following sum rules

\begin{eqnarray}
A^m_{N-1} + A^m_{N} + A^m_{N+1} & = & 1 \nonumber \\
(N-1) A^m_{N-1} + N A^m_{N} + (N+1) A^m_{N+1} & = & \sum_{l \ne m} n_l 
\nonumber \\
(N-1)^2 A^m_{N-1} + N^2 A^m_{N} + (N+1)^2 A^m_{N+1} & = & 
\sum_{l \ne m} n_l + <\hat{n} \hat{n}>_m, 
\end{eqnarray}
where $<\hat{n} \hat{n}>_m = \sum_{(l \ne k )\ne m} <\hat{n}_l \hat{n}_k>$. 
For the special case $N=0$ ($N=M-1$) one has $A^m_{N-1}=0$ ($A^m_{N+1}=0$) 
and only the first two Eqs. in (7) have to be considered.

This approximated expression for $G^{(at)}_m(\omega)$ is now fully
determined by the average charges $n_l$ and the two-body correlation
functions $<\hat{n}_l \hat{n}_k>$. As in the single-level case one can
define an atomic self-energy, $\Sigma^{(at)}_m = \omega -
\epsilon^{HF}_m - \left[G_m^{(at)}(\omega)\right]^{-1}$, which can be
written as the ratio of two polynomials in $\omega$
\begin{equation}
\Sigma^{(at)}_m  = \frac{a_m U^2  (\omega - \epsilon_m + i0^+) + b_m U^3}
{(\omega - \epsilon_m + i0^+)^2 +  
c_m U (\omega - \epsilon_m + i0^+) + d_m U^2} ,
\end{equation} 

\noindent
where $a_m = ({\cal N} - n_m) \left[1 - ({\cal N} - n_m) \right] +
<\hat{n} \hat{n}>_m$; $c_m = {\cal N} - n_m - 3N$; $d_m = <\hat{n}
\hat{n}>_m + 3N^2 - 1 - (3N-1)({\cal N} - n_m)$ and
$b_m = N^2(1 - N) - ({\cal N} - n_m) d_m$.

On the other hand, in the $U/\Gamma \rightarrow 0$ limit the self-energy
is accurately given by second order perturbation theory as in the single
level case. The second order self-energy $\Sigma^{(2)}_m$ now takes into
account the interaction of an electron on the dot level $m$ with
electron-hole pairs on each one of the other channels. 

For the interpolation one notices that both $\Sigma^{(2)}$ and
$\Sigma^{(at)}$ have the same functional form when extrapolated to the
corresponding opposite limit. The natural generalization of the ansazt 
in the single
level case now has the form of a continued fraction

\[ \Sigma_m(\omega) = \frac{\alpha_m \Sigma^{(2)}_m(\omega)}{1 - \beta_m
\Sigma^{(2)}_m(\omega) -  R_m(\omega)} \;\;,\;\;
R_m(\omega) =  \frac{\gamma_m (\Sigma^{(2)}_m(\omega))^2}{1 - \delta_m 
\Sigma^{(2)}_m(\omega)} \]
with coefficients $\alpha_m = U^2 a_m/\Delta_m$, $\beta_m = (\epsilon_m -
\tilde{\epsilon}_m + (b_m/a_m - c_m)U)/\Delta_m$, $\gamma_m = ((c_m -
b_m/a_m)b_m/a_m - d_m)U^2/\Delta_m^2$ and $\delta_m = (\epsilon_m -
\tilde{\epsilon}_m + Ub_m/a_m)/\Delta_m$, where $\Delta_m = U^2 \sum_{l \ne
m} \tilde{n}_l(1-\tilde{n}_l)$.                                    

As in the single-level case the final step is to determine the effective
levels self-consistently. In the multilevel case one has, in addition to
self-consistently determine the two-body correlations $<\hat{n} \hat{n}>_m$
that appear in the atomic self-energy. This can be done by means of the
relation 
\begin{equation}
\sum_{l \ne m} <\hat{n}_l \hat{n}_m> = -\frac{1}{\pi U}
\int_{-\infty}^{\infty} f(\omega)
\mbox{Im} \left[\left(\omega - \epsilon_m - \Gamma_m \right) 
G_m(\omega) \right] d\omega ,
\end{equation}
connecting the two-body correlations and the Green functions that can be
derived from the equation of motion for $G_m(\omega)$. This step turns
out to be essential in order to obtain the correct values of the charges
in the large $U$ limit. 

Finally, the temperature dependent dot linear 
conductance can be obtained using the expression \cite{MW}

\[ G = \frac{e^2}{h} \sum_m \frac{|\Gamma_{m,L}
\Gamma_{m,R}|}{(|\Gamma_{m,L}| + |\Gamma_{m,R}|)} \int_{-\infty}^{\infty} 
\left(\frac{\partial f}{\partial \omega}\right)
\mbox{Im} G^r_m(\omega) d\omega \]

\section{Results}

The multilevel formalism allows to study the importance of the
multilevel structure in the QD transport properties. 
For this purpose we have analyzed the $M=4$ case which corresponds to
two consecutive dot levels plus spin degeneracy. We have studied this
case as a function of the level separation $\Delta$. 

Figure 2 shows the dot conductance as a function of Fermi energy and
temperature for the cases $\Delta = 0$, 0.1 and 0.5 (in units of $U$). 

\begin{figure}[thbp]
\psfig{file=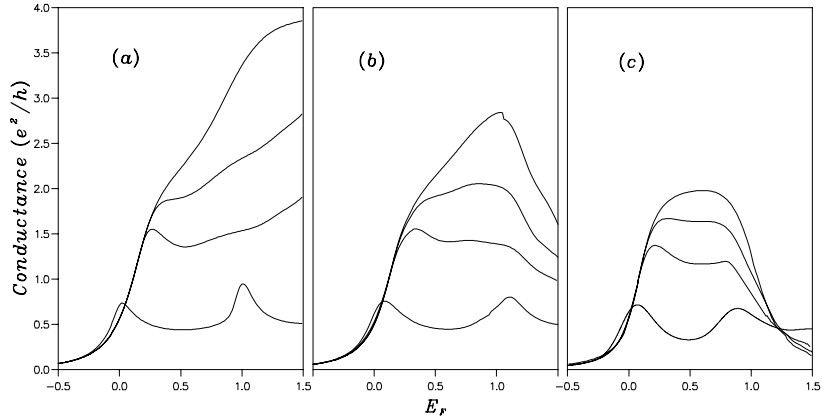,height=5cm}
\caption[]{Dot conductance as a function of Fermi energy for increasing
temperatures with $\Gamma_L = \Gamma_R = 0.075$ 
and $\Delta = 0$ (a), 0.1 (b) and 0.5 (c). The temperature values are
0.0005, 0.0025, 0.005 and 0.03 in units of $U$}
\end{figure}

This figure illustrates the transition from a two-fold degenerate
situation ($\Delta=0$), where the conductance reaches a maximum value $4
e^2/h$ for the half-filled case at zero temperature, to the case of well
separated dot levels, where the maximum conductance $2e^2/h$ is reached
for the quarter and three quarter filling case. The increase of conductance 
with decreasing temperature is due to the Kondo effect. While in
the case of well separated levels one observes only the Kondo effect due
to the spin-degeneracy of the individual levels, when $\Delta$ is small
compared to $\Gamma$ one can observe Kondo features involving the two
nearby dot levels. When the temperature is raised above the Kondo
temperature (which is around 0.005 for the parameters used in this figure)
one recovers the sequence of dot resonances at the charge degeneracy
points characteristic of the Coulomb blockade regime.

The Kondo effect should manifest also as a zero-bias anomaly in the dot
non-linear conductance. This anomaly is directly related to the
appearance of a narrow peak around the Fermi energy in the dot spectral
density. In cases where the splitting between dot levels is of the order
of $\Gamma$ we expect to have a zero-bias anomaly not only between dot
resonances corresponding to the same dot level but also in between
resonances corresponding to different levels. 
This feature is illustrated in Fig. 3
where we plot the density of states around $E_F$ for the same
three cases of Fig. 2 with $E_F=1.5$. 
The appearance of a zero-bias anomaly in between resonances
corresponding to different levels is a clear manifestation of the
multilevel structure of the QD which has
already been observed in recent experiments on semiconducting 
quantum dots  \cite{Weis}.

\begin{figure}[thbp]
\psfig{file=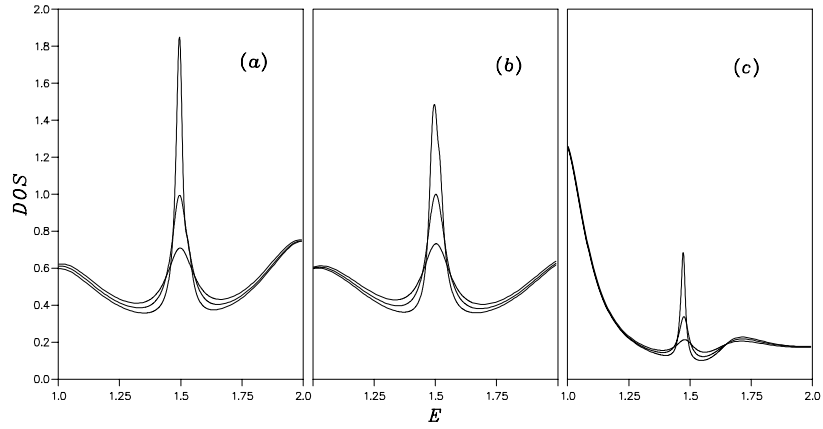,height=5cm}
\caption[]{Density of states around the Fermi energy for increasing
temperatures values (0.01, 0.02 and 0.03) for the same three cases
in Fig. 2 and $E_F=1.5$}
\end{figure}

This work has been funded by the Spanish CICyT under
contracts PB97-0028 and PB97-0044.

\end{document}